\documentclass[12pt]{article}
\usepackage[dvips]{graphics}

\def\beq   {\begin{equation}}
\def\eeq   {\end{equation}}
\def\beqd  {\begin{displaymath}}
\def\eeqd  {\end{displaymath}}
\def\beqaa {\begin{eqnarray}}
\def\eeqaa {\end{eqnarray}}

\def\noi {\noindent}

\def\ti  {\tilde}

\def\sf  {\ti f}

\def\st  {\ti t}
\def\sb  {\ti b}
\def\sg  {\ti g}
\def\sa  {\ti \tau}
\def\sn  {\ti \nu}

\def\nt  {\tilde\chi^0}
\def\ch  {\tilde\chi^\pm}
\def\chp {\tilde\chi^+}
\def\chm {\tilde\chi^-}

\def\a   {\alpha}
\def\b   {\beta}
\def\t   {\theta}

\def\tsa {\theta_{\sa}}

\def\sz{\ifmmode{\tilde{\chi}^0} \else{$\tilde{\chi}^0$} \fi}
\def\sw{\ifmmode{\tilde{\chi}} \else{$\tilde{\chi}$} \fi}

\newcommand{\lsim}{\;\raisebox{-0.9ex}{$\textstyle\stackrel{\textstyle<}
           {\sim}$}\;}

\begin{document}
\pagestyle{empty}

\vspace*{-1cm} 
\begin{flushright}
  TGU-29 \\
  UWThPh-2002-6 \\
  ZU-TH 7/02 \\
  hep-ph/0204071
\end{flushright}

\vspace*{1.4cm}

\begin{center}

{\Large {\bf
Impact of CP phases on the search for \\
sleptons \boldmath{$\tilde \tau$} and \boldmath{$\tilde \nu_\tau$}
}}\\

\vspace{10mm}

{\large 
A.~Bartl$^a$, K.~Hidaka$^b$, T.~Kernreiter$^a$ and W.~Porod$^c$}

\vspace{6mm}

\begin{tabular}{l}
$^a${\it Institut f\"ur Theoretische Physik, Universit\"at Wien, A-1090
Vienna, Austria}\\
$^b${\it Department of Physics, Tokyo Gakugei University, Koganei,
Tokyo 184--8501, Japan}\\
$^c${\it Institut f\"ur Theoretische Physik, Universit\"at Z\"urich, 
CH-8057 Z\"urich, Switzerland}
\end{tabular}

\end{center}

\vfill

\begin{abstract} 
We study the decays of the $\tau$-sleptons ($\sa_{1,2}$) and 
$\tau$-sneutrino ($\sn_\tau$) in the Minimal Supersymmetric 
Standard Model (MSSM) with complex parameters $A_\tau,\mu$ and $M_1$
(U(1) gaugino mass). We show that the effect of the CP phases of these 
parameters on the branching ratios of $\sa_{1,2}$ and $\sn_\tau$ 
decays can be quite strong in a large region of the MSSM parameter 
space. This could have an important impact on the search for $\sa_{1,2}$ 
and $\sn_\tau$ and the determination of the MSSM parameters at future 
colliders.
\end{abstract}

\newpage
\pagestyle{plain}
\setcounter{page}{2}

So far most phenomenological studies on supersymmetric (SUSY) particle 
searches have been performed in the Minimal Supersymmetric Standard Model 
(MSSM) \cite{ref1} with real SUSY parameters. 
Studies of the 3rd generation sfermions are particularly interesting 
because of the effects of the large Yukawa couplings. The lighter sfermion
mass eigenstates may be relatively light and they could be thoroughly 
studied at an $e^- e^+$ linear collider \cite{acco}. They could also be 
copiously produced in the decays of heavier SUSY particles. 
An analysis of the decays of 
the 3rd generation sleptons $\sa_2$ and $\sn_\tau$ in the MSSM with real 
parameters was performed in Ref.\cite{slepton}, and phenomenological studies of 
production and decays of the 3rd generation sfermions at future $e^- e^+$ 
colliders in Ref.\cite{rsf}. The assumption that all SUSY
parameters are real, however, may be too restrictive. The higgsino mass
parameter $\mu$ and the trilinear scalar coupling parameters $A_f$ of the
sfermions $\sf$ may be complex. In minimal Supergravity-type models the phase 
of $\mu$ ($\varphi_{\mu}$) turns out to be restricted by the experimental data 
on electron and neutron electric dipole moments (EDMs) to a range 
$|\varphi_{\mu}| \lsim 0.1-0.2$ for a universal scalar mass parameter 
$M_0 \lsim 400$GeV, while the phase of the universal trilinear scalar coupling 
parameter $A_0$ is correlated with $\varphi_{\mu}$, but otherwise unrestricted 
\cite{ref13'}. In more general models the phases of the parameters $A_f$ 
of the 3rd generation sfermions are not restricted at one-loop level by the 
EDM data. 
However, there may be restrictions at two-loop level \cite{Pilaftsis}. 
In a complete phenomenological analysis of production and decays of 
the SUSY particles one has to take into account that the $\mu$ and $A_f$ may 
be complex. 
Furthermore, explicit CP violation in the Higgs sector can be 
induced by loop effects involving CP-violating 
%
interactions of Higgs bosons to top and bottom squark ($\st$ and $\sb$) 
sector with complex parameters \cite{ref2,ref3}. It is found \cite{ref4} 
that such effects of the complex phases on the phenomenology of the Higgs 
boson search could be quite significant. In principle, the imaginary parts 
of the possible complex SUSY parameters involved could most directly and 
unambiguously be determined by measuring relevant CP-violating observables; 
e.g. such analyses in $\tau$-slepton ($\sa$) pair production in $e^+ e^-$ and 
$\mu^+ \mu^-$ colliders were performed in Ref.\cite{ref5}.\\
On the other hand, the CP-conserving observables also can depend on the 
phases of the complex parameters because in general the mass-eigenvalues 
and the couplings of the SUSY particles (sparticles) involved are functions 
of the underlying complex parameters. For example, the branching ratios of 
the Higgs boson decays depend strongly on the complex phases of the $\st$ 
and $\sb$ sectors \cite{ref5'}.

In this article we study the effects of the complex phases of the stau and 
gaugino-higgsino sectors on the decay branching ratios of the staus 
$\sa_{1,2}$ and $\tau$-sneutrino $\sn_\tau$ with $\sa_1$($\sa_2$) being the 
lighter (heavier) stau. 
We point out that these effects can be quite strong in a large region of 
the MSSM parameter space. This could have an important impact on the search 
for $\sa_{1,2}$ and $\sn_\tau$ and the determination of the MSSM parameters 
at future colliders.

First we summarize the MSSM parameters in our analysis. 
In the MSSM the stau sector is specified by the mass matrix in 
the basis $(\sa_L^{},\sa_R^{})$ \cite{ref7}
\begin{equation}
  {\cal M}^2_{\sa}= 
     \left( \begin{array}{cc} 
                m_{\sa_L}^2 & a_\tau^* m_\tau \\
                a_\tau m_\tau     & m_{\sa_R}^2
            \end{array} \right) 
                                                         \label{eq:a}
\end{equation}
with
\begin{eqnarray}
  m_{\sa_L}^2 &=& M_{\ti L}^2 
                  + m_Z^2\cos 2\beta\,(\sin^2\t_W - \frac{1}{2}) 
                  + m_\tau^2,                            \label{eq:b} \\
  m_{\sa_R}^2 &=& M_{\ti E}^2  
                  - m_Z^2 \cos 2\b\, \sin^2\t_W + m_\tau^2, 
                                                         \label{eq:c} \\[2mm]
  a_\tau m_\tau     &=& 
                     (A_\tau - \mu^* \tan\beta)\,m_\tau \,\,
                      = \,\, |a_\tau m_\tau| \, e^{i\varphi_{\sa}} \,\,
                      (-\pi < \varphi_{\sa} \leq \pi).
                                                          \label{eq:d}
\end{eqnarray}
$M_{\ti L,\ti E}$ and $A_\tau$ are soft SUSY--breaking 
parameters, $\mu$ is the higgsino mass parameter, 
and $\tan\b = v_2/v_1$ with $v_1$ $(v_2)$ being the vacuum 
expectation value of the Higgs field $H_1^0$ $(H_2^0)$. 
As the relative phase $\xi$ between $v_1$ and $v_2$ is irrelevant in our 
analysis, we adopt the $\xi = 0$ scheme \cite{ref3}. We take $A_\tau$ and 
$\mu$ as complex parameters: $A_\tau = |A_\tau| \, e^{i\varphi_{A_\tau}}$ 
and $\mu = |\mu| \, e^{i\varphi_{\mu}}$ with 
$-\pi < \varphi_{A_\tau,\mu} \leq \pi $. Diagonalizing the matrix 
(\ref{eq:a}) one gets the mass eigenstates $\sa_1$ and $\sa_2$
\begin{equation}
     \left( \begin{array}{c} 
                \sa_1 \\
                \sa_2
            \end{array} \right) = R^{\sa} 
     \left( \begin{array}{c} 
                \sa_L \\
                \sa_R
            \end{array} \right) = 
     \left( \begin{array}{cc} 
                e^{i\varphi_{\sa}} \cos\tsa & \sin\tsa \\
                -\sin\tsa  & e^{-i\varphi_{\sa}} \cos\tsa
            \end{array} \right) 
     \left( \begin{array}{c} 
                \sa_L \\
                \sa_R
            \end{array} \right)
                                                         \label{eq:e}
\end{equation}
with the masses $m_{\sa_1}$ and $m_{\sa_2}$ ($m_{\sa_1} < m_{\sa_2}$), 
and the mixing angle $\tsa$
\begin{eqnarray}
  m_{\sa_{1,2}}^2 &=& \frac{1}{2}(m_{\sa_L}^2 + m_{\sa_R}^2 
    \mp \sqrt{ (m_{\sa_L}^2 - m_{\sa_R}^2)^2 + 4|a_\tau m_\tau|^2 }), 
                                                           \label{eq:f} \\
  \tsa &=& \tan^{-1}(|a_\tau m_\tau|/(m_{\sa_1}^2 - m_{\sa_R}^2)) \quad 
                                         (-\pi/2 \leq \tsa \leq 0).
                                                           \label{eq:g} 
\end{eqnarray}
The $\sa_L - \sa_R$ mixing is large if $|m_{\sa_L}^2 - m_{\sa_R}^2| 
\lsim |a_\tau m_\tau|$, which may be the case for large $\tan\b$ and 
$|\mu|$. From Eqs.(\ref{eq:f}) and (\ref{eq:g}) we see that 
$m_{\sa_{1,2}}^2$ and $\tsa$ depend on the phases only through a term 
$\cos(\varphi_{A_\tau} + \varphi_{\mu})$. This phase dependence is 
strongest if $|A_\tau| \simeq |\mu| \tan\b$. The mass of $\sn_\tau$ is 
given by 
\begin{equation}
  m_{\sn_\tau}^2 = M_{\ti L}^2 + \frac{1}{2} m_Z^2 \cos 2\b.
                                                             \label{eq:h}
\end{equation}

The properties of charginos $\ch_i$ ($i=1,2$; $m_{\ch_1}<m_{\ch_2}$) and 
neutralinos $\nt_j$ ($j=1,...,4$; $m_{\nt_1}< ... < m_{\nt_4}$) are 
determined by the parameters $M_2, M_1, \mu$ and $\tan\b$, where $M_2$ 
and $M_1$ are the SU(2) and U(1) gaugino masses, respectively. We assume 
that $M_2$ and the gluino mass $m_{\sg}$ are real and that $M_1$ is 
complex: $M_1 = |M_1| e^{i\varphi_1} \, (-\pi < \varphi_1 \leq \pi)$. 
Inspired by the gaugino mass unification we take 
$|M_1| = (5/3) \tan^2\t_W M_2$ and $m_{\sg} = (\a_s(m_{\sg})/\a_2) M_2$. 
In the MSSM Higgs sector with explicit CP violation the 
mass-eigenvalues and couplings of the neutral and charged Higgs bosons 
$H_1^0, H_2^0, H_3^0$ $(m_{H_1^0} < m_{H_2^0} < m_{H_3^0})$ and $H^\pm$, 
including Yukawa and QCD radiative corrections, are fixed by 
$m_{H^+},\tan\b,\mu, m_t,m_b,M_{\ti Q},M_{\ti U},M_{\ti D},A_t,A_b,
|M_1|,M_2$, and $m_{\sg}$ \cite{ref3}. Here $M_{\ti Q,\ti U,\ti D}$ and 
$A_{t,b}$ are the soft SUSY-breaking parameters in the $\st$ and $\sb$ 
sectors, and $A_{t,b}$ are in general also complex: 
$A_{t,b} = |A_{t,b}| e^{i\varphi_{A_{t,b}}} \, 
(-\pi < \varphi_{A_{t,b}} \leq \pi)$. The neutral Higgs mass eigenstates 
$H_1^0, H_2^0$ and $H_3^0$ are mixtures of CP-even and CP-odd states 
($\phi_{1,2}$ and $a$) due to the explicit CP violation in the Higgs sector. 
For the radiatively corrected masses and mixings of the Higgs bosons we use 
the formulae of Ref.\cite{ref3}. We treat 
$M_{ \{ \ti L,\ti E,\ti Q,\ti U,\ti D \} }$ and $A_{ \{ \tau,t,b \} }$ as 
free parameters since the ratios among them are highly model-dependent. 

Here we list possible important decay modes of $\sa_{1,2}$ and $\sn_\tau$: 
\beqaa
  \sa_1 & \to & \tau \, \nt_i \, , \, \nu_\tau \, \chm_j 
                                                           \label{eq:i} \\
  \sa_2 & \to & \tau \, \nt_i \, , \, \nu_\tau \, \chm_j \, , \, 
                \sa_1 Z^0 \, , \, \sn_\tau W^- \, , \, \sa_1 H_k^0 \, , \, 
                \sn_\tau H^- 
                                                           \label{eq:j} \\
  \sn_\tau & \to & \nu_\tau \, \nt_i \, , \, \tau \, \chp_j \, , \, 
                   \sa_1 W^+ \, , \, \sa_1 H^+. 
                                                           \label{eq:k} 
\eeqaa
The decays into a gauge or Higgs boson in (\ref{eq:j}) and (\ref{eq:k}) 
are possible in case the mass splitting between the sleptons is sufficiently 
large \cite{slepton}. The explicit expressions of the widths of the decays 
(\ref{eq:i})-(\ref{eq:k}) in case of real SUSY parameters are given in 
\cite{ref8}. Those for complex parameters can be obtained by using the 
corresponding masses and couplings (mixings) from 
Refs.\cite{ref3,ref7} and will be presented elsewhere \cite{ref9}.

The phase dependence of the widths stems from that of the involved 
mass-eigenvalues, mixings and couplings among the interaction-eigenfields. 
Here we summarize the latter phase dependence:
\renewcommand{\labelenumi}{(\Roman{enumi})} 
\begin{enumerate}
  \item $\sa_i$ sector:
    \begin{enumerate}
      \item $m_{\sa_{1,2}}$ are insensitive to the phases 
            $(\varphi_{A_\tau}, \varphi_{\mu})$ 
            if the $\sa$-mixing term $|a_\tau m_\tau| \ll 
            m_{\sa_L}^2 + m_{\sa_R}^2$. In most cases this is naturally 
            fulfilled, because $m_\tau$ is small. 
      \item The $\sa$-mixing angle $\tsa$ (given by 
            $\tan 2\tsa = 2|a_\tau m_\tau|/(m_{\sa_L}^2 - m_{\sa_R}^2)$ ) 
            is sensitive to $(\varphi_{A_\tau}, \varphi_{\mu})$ (via 
            $\cos(\varphi_{A_\tau} + \varphi_{\mu})$ ) if and only if 
            $|m_{\sa_L}^2 - m_{\sa_R}^2|$ is small {\em and} 
            $|A_\tau| \sim |\mu|\tan\b$. Here note that 
            $|m_{\sa_L}^2 - m_{\sa_R}^2| \simeq |M_{\ti L}^2 - M_{\ti E}^2| 
            \simeq |m_{\sa_1}^2 - m_{\sa_2}^2|$ due to the smallness of 
            $|a_\tau m_\tau|$.
    \end{enumerate}
  \item $\nt_i$ and $\ch_j$ sectors:
    \begin{enumerate}
      \item $m_{\nt_i}$ (i=1,...,4) and the $\nt$-mixing matrix are sensitive 
            $[$insensitive$]$ to the phases $(\varphi_1, \varphi_{\mu})$ 
            for small $[$large$]$ $\tan\b$.
      \item $m_{\ch_{1,2}}$ and the $\ch$-mixing matrices are sensitive 
            $[$insensitive$]$ to $\varphi_{\mu}$ for small $[$large$]$ 
            $\tan\b$.
    \end{enumerate}
  \item $\sn_\tau$ and $H^\pm$ sectors:\\
        These sectors are independent of the phases. 
  \item $H_i^0$ sector:
    \begin{enumerate}
      \item $m_{H_i^0}$ (i=1,2,3) are sensitive $[$insensitive$]$ to the 
            phase sums $\varphi_{A_{t,b}} + \varphi_{\mu}$ for small 
            $[$large$]$ $\tan\b$ \cite{ref3}.
      \item In general the $H_i^0$-mixing matrix (a real orthogonal 
            $3 \times 3$ matrix $O_{ij}$) is sensitive to 
            $\varphi_{A_{t,b}} + \varphi_{\mu}$ for any $\tan\b$ \cite{ref3}.
    \end{enumerate}
  \item The couplings among the interaction-eigenfields:
    \begin{enumerate}
      \item For the decays into fermions and gauge bosons in 
            Eqs.(\ref{eq:i})-(\ref{eq:k}), they are gauge couplings and$/$or 
            tau Yukawa coupling ($h_\tau$), and are independent of 
            the phases. 
      \item For the decays into Higgs bosons in Eqs.(\ref{eq:j}) and 
            (\ref{eq:k}), the slepton-``chirality" flip (nonflip) couplings are 
            dependent on (independent of) the phases $\varphi_{A_\tau}$ and 
            $\varphi_{\mu}$;
            \beqaa
              C(\sn_\tau^\dagger \sa_R H^+) & \sim & \cos\b \, h_\tau 
                                              (A_\tau^* \tan\b + \mu) 
                                                           \label{eq:l} \\
              C(\sa_L^\dagger \sa_R \phi_1) & \sim & h_\tau A_\tau^* 
                                                           \label{eq:m} \\
              C(\sa_L^\dagger \sa_R \phi_2) & \sim & h_\tau \mu 
                                                           \label{eq:n} \\
              C(\sa_L^\dagger \sa_R a) & \sim & \cos\b \, h_\tau 
                                                (A_\tau^* \tan\b + \mu) 
                                                           \label{eq:o} \\
              C(\sn_\tau^\dagger \sa_L H^+) & \sim & 
                           C(\sa_L^\dagger \sa_L \phi_{1,2}) \, \sim \,
                           C(\sa_R^\dagger \sa_R \phi_{1,2}) \, \sim \, g m_W 
                                                           \label{eq:p} 
            \eeqaa
            with 
            \beq
              h_\tau = g m_\tau / (\sqrt{2} m_W \cos\b). 
                                                           \label{eq:q} 
            \eeq
            Here $\phi_i = O_{ij} H_j^0$ (i=1,2) and $a = O_{3j} H_j^0$ 
            are the CP-even and CP-odd neutral Higgs bosons, respectively 
            \cite{ref3}. 
    \end{enumerate}
\end{enumerate}
From the facts (I)-(V) the widths (and hence the branching ratios) of the 
decays (\ref{eq:i})-(\ref{eq:k}) are expected to be sensitive to the phases 
($\varphi_{A_\tau}$, $\varphi_{\mu}$, $\varphi_1$, 
$\varphi_{A_{t,b}} + \varphi_{\mu}$) in a large region of the MSSM 
parameter space. 

Now we turn to the numerical analysis of the $\sa_{1,2}$ and $\sn_\tau$ decay 
branching ratios. We calculate the widths of all possibly important two-body 
decay modes of Eqs.(\ref{eq:i})-(\ref{eq:k}). Three-body decays are negligible 
in this study. We take  $m_\tau = 1.78$ GeV, $m_t=175$ GeV, $m_b=5$ GeV, 
$m_Z=91.2$ GeV, $\sin^2\t_W =0.23$, $m_W = m_Z \cos\t_W$, 
$\a(m_Z)=1/129$, and $\alpha_s(m_Z)=0.12$ (with 
$\alpha_s(Q)=12\pi/((33-2n_f)\ln(Q^2/\Lambda_{n_f}^2))$ for the determination 
of $m_{\sg}(=(\alpha_s(m_{\sg})/\alpha_2)M_2)$, $n_f$ being the number 
of quark flavors), where $m_{t,b}$ are pole masses of t and b quarks. 
In order not to vary too many parameters we fix $m_{\sa_1}=240$GeV, 
$m_{H^+}=180$GeV, $M_{\ti Q}=M_{\ti U}=M_{\ti D}=|A_t|=|A_b|=1000$GeV, and 
$\varphi_{A_t} = \varphi_{A_b} = 0$ for simplicity. In our numerical study 
we take $\tan\b$, $M_2$, $m_{\sa_2}$, $|A_\tau|$, $|\mu|$, $\varphi_{A_\tau}$, 
$\varphi_{\mu}$ and $\varphi_1$ as input parameters. Note that for a given 
set of the input parameters we have two solutions for 
($M_{\ti L}$, $M_{\ti E}$) corresponding to the two cases 
$m_{\sa_L} \geq m_{\sa_R}$ and $m_{\sa_L} < m_{\sa_R}$. In the plots we 
impose the following conditions in order to respect experimental and 
theoretical constraints:
\renewcommand{\labelenumi}{(\roman{enumi})} 
\begin{enumerate}
  \item $m_{\ch_1} > 103$ GeV, $m_{\nt_1} > 50$ GeV,
        $m_{\sa_1,\st_1,\sb_1} > 100$ GeV, 
        $m_{\sa_1,\st_1,\sb_1} > m_{\nt_1}$, \\
        $m_{H_1^0} > $ 110 GeV,  
  \item $|A_\tau|^2 < 3\,(M_{\ti L}^2 + M_{\ti E}^2 + m_1^2)$, 
        $|A_t|^2 < 3\,(M_{\ti Q}^2 + M_{\ti U}^2 + m_2^2)$, and 
        $|A_b|^2 < 3\,(M_{\ti Q}^2 + M_{\ti D}^2 + m_1^2)$, where 
        $m_1^2=(m_{H^+}^2 + m_Z^2 \sin^2\t_W)\sin^2\b-\frac{1}{2}\,m_Z^2$ and 
        $m_2^2=(m_{H^+}^2 + m_Z^2 \sin^2\t_W)\cos^2\b-\frac{1}{2}\,m_Z^2$, 
  \item $\Delta\rho\,(\st \!-\! \sb) < 0.0012$ 
        \cite{ref10} using the formula of \cite{ref11}.
\end{enumerate}
Condition (i) is imposed to satisfy the experimental mass bounds from LEP 
\cite{ref12}. (ii) is the approximate necessary condition for the tree-level 
vacuum stability \cite{ref13}. (iii) constrains $\mu$ and $\tan\b$ (in the 
squark sector). We do not impose the $b \to s \gamma$ constraint \cite{Tata} 
since it strongly depends on the details of the flavour structures of the 
squarks, including the generation-mixing.


\noi
In general the experimental upper limits on the electron and neutron electric 
dipole moments (EDMs) strongly constrain the SUSY CP phases \cite{ref13'}. 
One interesting possibility for evading these constraints is to invoke large 
masses (much above the TeV scale) for the first two generations of the 
sfermions \cite{ref14}, keeping the third generation sfermions relatively 
light ($\lsim$1 TeV). In such a scenario ($\varphi_1$, $\varphi_{\mu}$) and 
the CP phases in the third generation ($\varphi_{A_\tau}$, $\varphi_{A_t}$, 
$\varphi_{A_b}$) are practically unconstrained \cite{ref14}. We take this 
scenario. 
The deviation of the recent data on the muon g-2 from the Standard 
Model prediction is no longer significant \cite{muon g-2}, which allows of 
our scenario. 
We have checked that the electron and neutron EDM constraints at two-loop 
level \cite{Pilaftsis} are fulfilled in the numerical examples studied 
in this article. 


In Fig.1 we plot the contours of the branching ratios of the $\sa_1$ decays 
$B(\sa_1 \to \tau \nt_1)$ and $B(\sa_1 \to \nu_\tau \chm_1)$ in the 
$|A_\tau|-|\mu|$ plane for $\tan\b=3$, $M_2=200$GeV, $m_{\sa_1}=240$GeV, 
$m_{\sa_2}=250$GeV, $\varphi_1=0$, and ($\varphi_{A_\tau}$, $\varphi_{\mu}$) = 
(0,0), ($\pi/2$,0), (0,$\pi/2$) in the case $m_{\sa_L} < m_{\sa_R}$; in the case 
$m_{\sa_L} \geq m_{\sa_R}$ we have obtained similar results. As expected, 
these branching ratios are very sensitive to the phases $\varphi_{A_\tau}$ 
and $\varphi_{\mu}$ in a sizable region of the $|A_\tau|-|\mu|$ plane. 
As can be seen from item (I), in this case the $\sa$-mixing 
angle $\tsa$ is sensitive to cos$(\varphi_{A_\tau} + \varphi_{\mu})$ for 
$|A_\tau| \sim 3|\mu|$, which is the main cause for the difference between 
the case of ($\varphi_{A_\tau}$, $\varphi_{\mu}$) = (0,0) and those of 
($\pi/2$,0) and (0,$\pi/2$). Furthermore from item (II) one sees that the 
masses and mixing-matrices of the $\nt_i$ and $\ch_j$ are sensitive to 
$\varphi_{\mu}$, which is the main reason for the difference between the 
case of ($\varphi_{A_\tau}$, $\varphi_{\mu}$) = (0,$\pi/2$) and those of 
(0,0) and ($\pi/2$,0). Here note also item (V)-(a). 

In Fig.2 we plot the contours of the $\sa_1$ decay branching ratios 
$B(\sa_1 \to \tau \nt_1)$, $B(\sa_1 \to \tau \nt_2)$, and 
$B(\sa_1 \to \nu_\tau \chm_1)$ in the $\varphi_{A_\tau}-\varphi_{\mu}$ 
plane for $\tan\b=3$, $M_2=200$GeV, $m_{\sa_1}=240$GeV, $m_{\sa_2}=255$GeV, 
$|A_\tau|=600$GeV, $|\mu|=350$GeV, and $\varphi_1=0$ in the case 
$m_{\sa_L} < m_{\sa_R}$; in the case $m_{\sa_L} \geq m_{\sa_R}$ we have 
obtained similar results. 
One sees that these branching ratios depend on the CP phases 
$\varphi_{A_\tau}$ and $\varphi_{\mu}$ quite strongly, as expected 
from items (I) and (II). 

In Fig.3 we show the $\varphi_{\mu}$ dependence of the $\sa_1$ decay 
branching ratios for $\tan\b=3,30$ (with $\varphi_1=0$) and 
$\varphi_1=0,\pi/2$ (with $\tan\b=3$) with $M_2=300$GeV, $m_{\sa_2}=500$GeV, 
$|A_\tau|=600$GeV, $|\mu|=200$GeV, and $\varphi_{A_\tau}=0$ in the case 
$m_{\sa_L} < m_{\sa_R}$; in the case $m_{\sa_L} \geq m_{\sa_R}$ we have 
obtained similar results. From Fig.3a we see that the $\sa_1$ decay 
branching ratios are sensitive (insensitive) to $\varphi_{\mu}$ for small 
(large) $\tan\b$ as is expected from items (I), (II), and (V)-(a). 
In general the $\sa_1$ decay branching ratios are insensitive to all of the 
phases ($\varphi_{A_\tau}$, $\varphi_{\mu}$, $\varphi_1$) for large $\tan\b$ 
according to items (I), (II) and (V)-(a). 
From Fig.3b we find that they are sensitive to $\varphi_{\mu}$ and 
$\varphi_1$ for small $\tan\b$(=3) where the masses and mixings of $\nt_i$ 
$[\ch_j]$ are sensitive to ($\varphi_{\mu}$, $\varphi_1$) $[\varphi_{\mu}]$ 
(though $\tsa$ is insensitive to $\varphi_{\mu}$ in this case) as seen from 
items (I) and (II). 

In Fig.4 we show the $\varphi_{A_\tau}$ dependence of the $\sa_2$ decay 
branching ratios for $\varphi_\mu=0$ and $\pi/2$ in the two cases of small 
(a) and large (b,c) mass difference $|m_{\sa_1} - m_{\sa_2}|$. 
For the former case we take ($m_{\sa_1}$(GeV), $m_{\sa_2}$(GeV), 
$\tan\b$, $M_2$(GeV), $|A_\tau|$(GeV), $|\mu|$(GeV), $\varphi_1$) = 
(240, 260, 6, 500, 600, 150, 0) with $m_{\sa_L} \geq m_{\sa_R}$, whereas 
for the latter case we take ($m_{\sa_1}$(GeV), $m_{\sa_2}$(GeV), 
$\tan\b$, $M_2$(GeV), $|A_\tau|$(GeV), $|\mu|$(GeV), $\varphi_1$) = 
(240, 500, 30, 400, 900, 800, 0) with $m_{\sa_L} < m_{\sa_R}$. 
%
%
In Fig.4a (where $|m_{\sa_1} - m_{\sa_2}|$ and $|m_{\sn_\tau} - m_{\sa_2}|$ 
are so small that the bosonic decays in Eq.(\ref{eq:j}) are kinematically 
forbidden) we see that the $\sa_2$ decay branching ratios are very sensitive 
to $\varphi_{A_\tau}$ and $\varphi_{\mu}$ as expected from items (I) 
and (II). In Figs.4b and 4c (where $|m_{\sa_1} - m_{\sa_2}| 
(\sim |m_{\sn_\tau} - m_{\sa_2}|)$ is so large that the bosonic decays in 
Eq.(\ref{eq:j}) also are allowed) we find that the branching ratios of the 
Higgs boson modes are rather sensitive to $\varphi_{A_\tau}$ and 
$\varphi_{\mu}$ as expected from items (IV)-(b) and (V)-(b); here note 
that in this case $(\sa_1, \sa_2) \sim (\sa_L, \sa_R)$ due to the smallness 
of the $\sa_L$-$\sa_R$ mixing term and hence that the bosonic decays of 
$\sa_2$ are basically the decays of $\sa_R$ into $(\sa_L, \sn_\tau)$. 

In Fig.5 we show the $\varphi_{\mu}$ dependence of the $\sn_\tau$ decay 
branching ratios for $\varphi_1=0$ and $\pi/2$ in the case of $\tan\b=3$, 
$M_2=500$GeV, $m_{\sa_2}=500$GeV, $|A_\tau|=600$GeV, $|\mu|=150$GeV, 
$\varphi_{A_\tau}=0$, and $m_{\sa_L} < m_{\sa_R}$ (for which 
$m_{\sn_\tau} \sim m_{\sa_L} \sim m_{\sa_1}$ and hence the bosonic decays 
in Eq.(\ref{eq:k}) are kinematically forbidden). We see that the $\sn_\tau$ 
decay branching ratios are quite sensitive to $\varphi_{\mu}$ and 
$\varphi_1$ as expected from item (II). 
In general the $\sn_\tau$ decay branching ratios are insensitive to all of the 
phases ($\varphi_{A_\tau}$, $\varphi_{\mu}$, $\varphi_1$) for large $\tan\b$ 
as can be seen from items (I)$\sim$(III) and (V). Furthermore, they tend to be 
insensitive to $\varphi_{A_\tau}$ for any $\tan\b$ as can be seen from 
items (I)$\sim$(III) and (V), except for some special cases such as the case 
with $m_{\sa_L} > m_{\sa_R}$, $|m_{\sa_1} - m_{\sa_2}| > m_{H^+}$, small $\tan\b$,
and $|\mu| \ll m_{\sn_\tau} (\sim m_{\sa_2}) < |M_{1,2}|$, where 
$m_{\sn_\tau} > m_{\sa_1} + m_{H^+}$, 
$m_{ \{ \nt_{1,2}, \ch_1 \} } < m_{\sn_\tau} 
< m_{ \{ \nt_{3,4}, \ch_2 \} }$, $\sa_1 \sim \sa_R$, and 
$(\nt_{1,2}, \ch_1) [(\nt_{3,4}, \ch_2)]$ 
are higgsino-like [gaugino-like]; in this case only the $\tau \chp_1$ and 
$\sa_1 H^+$ modes dominate the $\sn_\tau$ decay, and 
the coupling $C(\sn_\tau \sa_1^\dagger H^-)  \sim C(\sn_\tau \sa_R^\dagger H^-) 
\sim \cos\b \, h_\tau (A_\tau \tan\b + \mu^*)$ can be rather sensitive to 
$\varphi_{A_\tau}$ (though the $\tsa$ is insensitive to $\varphi_{A_\tau}$), 
which results in significant $\varphi_{A_\tau}$-dependence of the $\sn_\tau$ 
decay branching ratios. We have checked that this can be easily realized indeed. 


As for the $\tau$-lepton EDM ($d_\tau$), we have checked that in the MSSM 
parameter region considered here the predicted range of it is well below 
the current experimental limit $(|d_\tau| < 3.1 \times 10^{-16} e \, cm)$ 
\cite{Particle Data} and most likely also below the expected sensitivity 
of future experiments to measure this EDM: we find that in the parameter region 
considered here $|d_\tau| \lsim 10^{-20} e \, cm$, which is obtained by using 
the corresponding formulas in Ref. \cite{Bartl EDM} with $m_e$ replaced by 
$m_\tau$.  

The CP phases can significantly affect not only CP-violating observables 
such as the lepton EDM but also CP-conserving quantities such as the 
branching ratios of the $\sa_{1,2}$ and $\sn_\tau$ decays. Hence the 
possible sizable phases could have important consequences for the 
determination of the fundamental MSSM parameters by measurements of 
CP-conserving observables from which they are extracted. 
%

In conclusion, we have shown that the effect of the CP phases of the complex 
parameters $A_\tau$, $\mu$ and $M_1$ on the branching ratios of the 
$\sa_{1,2}$ and $\sn_\tau$ decays can be quite strong in a large region of 
the MSSM parameter space. This could have an important impact 
on the search for $\sa_{1,2}$ and $\sn_\tau$ and the determination of the 
MSSM parameters at future colliders. 
%

\section*{Acknowledgements}
The authors thank A. Pilaftsis and C.E.M. Wagner for useful correspondence. 
This work was supported by the `Fonds zur F\"orderung der wissenschaftlichen 
Forschung' of Austria, Project No. P13139-PHY, and by the European Community's 
Human Potential Programme under contract HPRN-CT-2000-00149. 
W.P.~is supported by `Fonds zur F\"orderung der wissenschaftlichen 
Forschung' of Austria, Erwin Schr\"odinger fellowship Nr. J2095, and 
partly by the `Schweizer Nationalfonds'. 


\newpage



\begin{flushleft}
{\Large \bf Figure Captions} \\
\end{flushleft}

\noi
{\bf Figure 1}: 
Contours of the $\sa_1$ decay branching ratios $B(\sa_1 \to \tau \nt_1)$ 
(a,b,c) and $B(\sa_1 \to \nu_\tau \chm_1)$ (d,e,f) in the $|A_\tau|-|\mu|$ 
plane for $\tan\b=3$, $M_2=200$GeV, $m_{\sa_1}=240$GeV, $m_{\sa_2}=250$GeV, 
$\varphi_1=0$, and ($\varphi_{A_\tau}$, $\varphi_{\mu}$) = (0,0) (a,d), 
($\pi/2$,0) (b,e), and (0,$\pi/2$) (c,f) in the case $m_{\sa_L} < m_{\sa_R}$. 
The blank areas are excluded by the conditions (i) to (iii) given in the text 
and the inequality $(m_{\sa_L}^2 - m_{\sa_R}^2)^2 = 
(m_{\sa_1}^2 - m_{\sa_2}^2)^2 - (2|a_\tau m_\tau|)^2 \geq 0$. 
The excluded region is different in each plot mainly due to the fact 
that the inequality depends on the phases ($\varphi_{A_\tau}$, $\varphi_{\mu}$). 
%
%
%

\noi 
{\bf Figure 2}: 
Contours of the $\sa_1$ decay branching ratios $B(\sa_1 \to \tau \nt_1)$ (a), 
$B(\sa_1 \to \tau \nt_2)$ (b), and $B(\sa_1 \to \nu_\tau \chm_1)$ (c) in the 
$\varphi_{A_\tau}-\varphi_{\mu}$ plane for $\tan\b=3$, $M_2=200$GeV, 
$m_{\sa_1}=240$GeV, $m_{\sa_2}=255$GeV, $|A_\tau|=600$GeV,
$|\mu|=350$GeV, and $\varphi_1=0$ in the case $m_{\sa_L} < m_{\sa_R}$. 

\noi 
{\bf Figure 3}: 
$\varphi_{\mu}$ dependence of the $\sa_1$ decay branching ratios for 
$\tan\b=3,30$ (with $\varphi_1=0$) (a) and $\varphi_1=0,\pi/2$ 
(with $\tan\b=3$) (b) with $M_2=300$GeV, $m_{\sa_1}=240$GeV, $m_{\sa_2}=500$GeV, 
$|A_\tau|=600$GeV, $|\mu|=200$GeV, and $\varphi_{A_\tau}=0$ 
in the case $m_{\sa_L} < m_{\sa_R}$. 
In Fig.a the solid and dashed lines are for $\tan\b$ = 3 and 30, respectively. 
In Fig.b the solid and dashed lines are for $\varphi_1$ = 0 and $\pi/2$, 
respectively. 

\noi 
{\bf Figure 4}: 
$\varphi_{A_\tau}$ dependence of the $\sa_2$ decay branching ratios for 
$\varphi_\mu=0$ (solid lines) and $\pi/2$ (dashed lines) in two cases: 
(a) small $\sa$-mass-splitting case with $m_{\sa_1} = 240$GeV, 
$m_{\sa_2} = 260$GeV, $\tan\b = 6$, $M_2 = 500$GeV, $|A_\tau|=600$GeV, 
$|\mu|=150$GeV, $\varphi_1=0$, and $m_{\sa_L} \geq m_{\sa_R}$ (for which 
$m_{\sn_\tau} \sim m_{\sa_2}$), and (b,c) large $\sa$-mass-splitting case 
with $m_{\sa_1}=240$GeV, $m_{\sa_2}=500$GeV, $\tan\b=30$, $M_2=400$GeV, 
$|A_\tau|=900$GeV, $|\mu|=800$GeV, $\varphi_1=0$, and $m_{\sa_L} < m_{\sa_R}$ 
(for which $m_{\sn_\tau} \sim m_{\sa_1}$). In the latter case (Figs.b and c) 
we have $(m_{H_1^0},m_{H_2^0},m_{H_3^0})=(117.4,159.2,159.5)$(GeV) and 
$(117.7,150.6,151.2)$(GeV) for $\varphi_{\mu}=0 \; \mbox{and} \; \pi/2$, 
respectively. 
The branching ratios $B(\sa_2 \to \sa_1 Z^0) \, (\sim 10 \%)$ and 
$B(\sa_2 \to \sn_\tau W^-) \, (\sim 20 \%)$, which are rather insensitive to  
$\varphi_{A_\tau}$, are not shown in Figs.b and c. 

\noi 
{\bf Figure 5}: 
$\varphi_{\mu}$ dependence of the $\sn_\tau$ decay branching ratios for 
$\varphi_1=0$ (solid lines) and $\pi/2$ (dashed lines) with $\tan\b=3$, 
$M_2=500$GeV, $m_{\sa_1}=240$GeV, $m_{\sa_2}=500$GeV, $|A_\tau|=600$GeV, 
$|\mu|=150$GeV, $\varphi_{A_\tau}=0$, and $m_{\sa_L} < m_{\sa_R}$ 
(for which $m_{\sn_\tau} \sim m_{\sa_1}$). 

\newpage
%
%
%
\begin{figure}[!htb] 
\begin{center}
\scalebox{0.66}[0.66]{\includegraphics{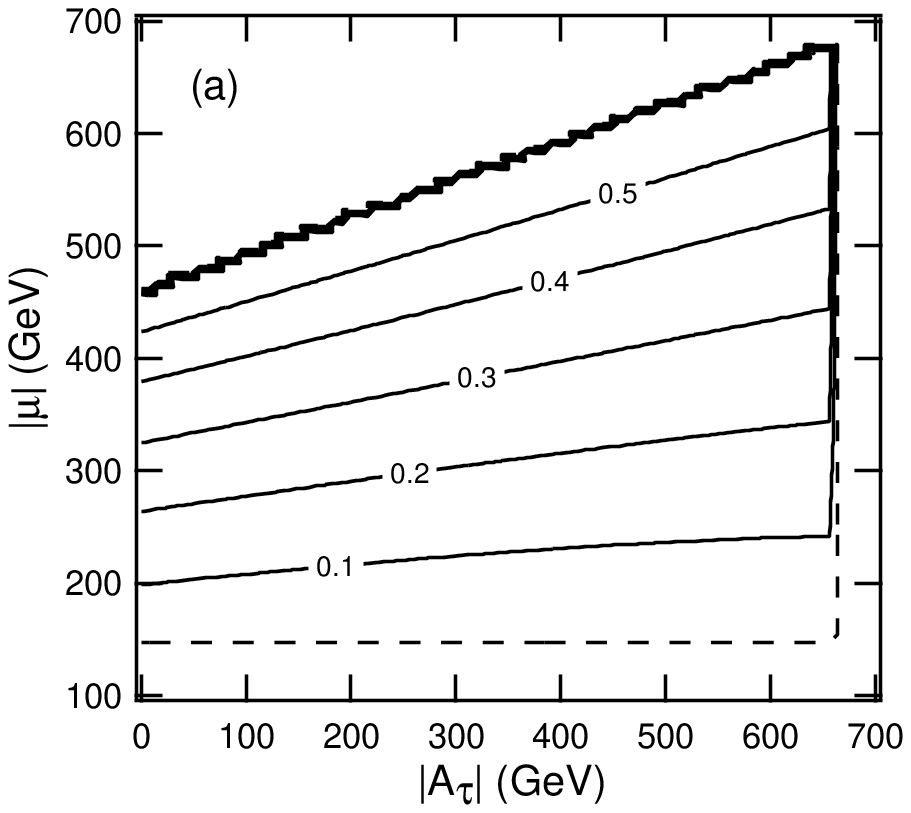}} \hspace{-7.5mm} 
\scalebox{0.66}[0.66]{\includegraphics{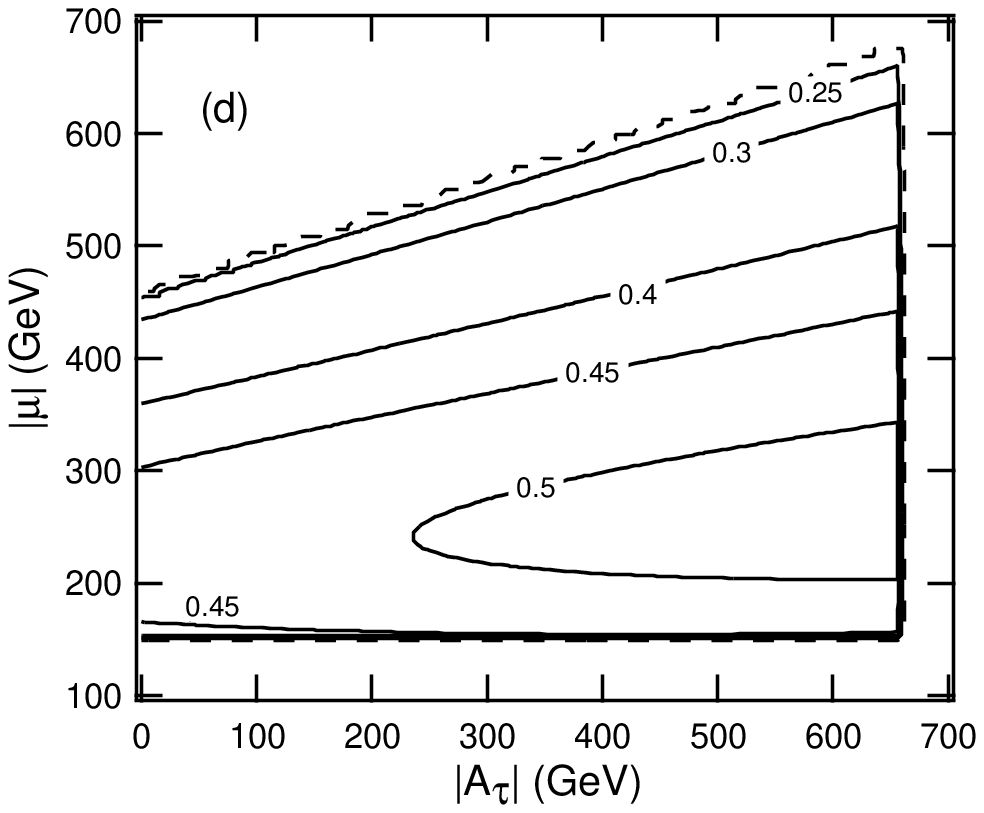}} \\ 
\scalebox{0.66}[0.66]{\includegraphics{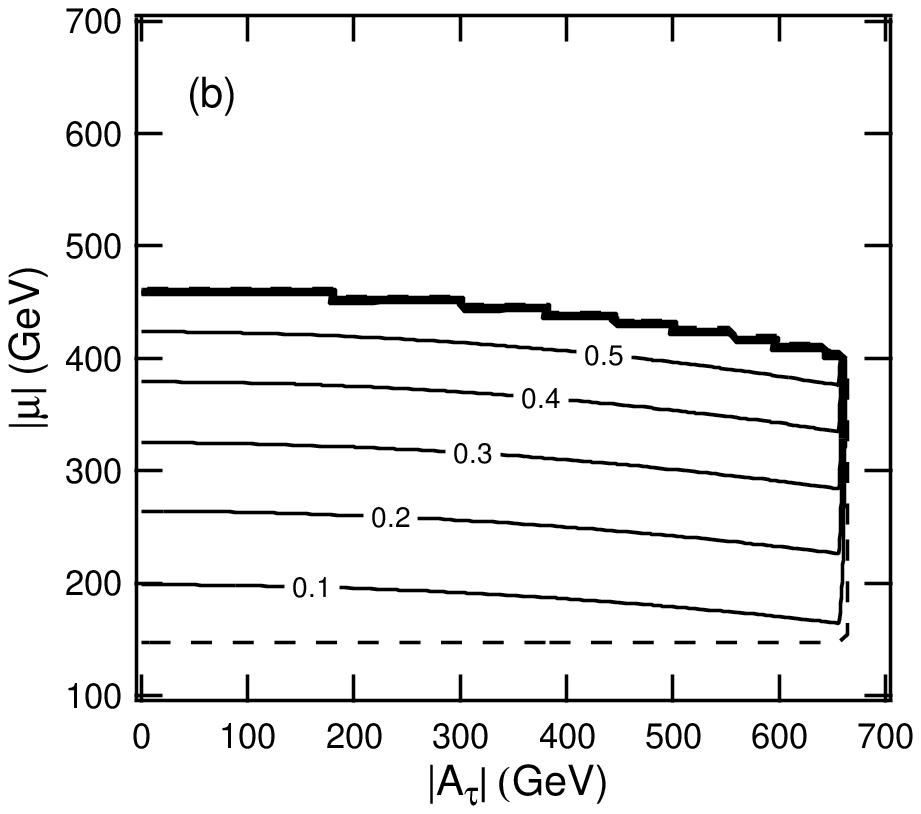}} \hspace{-7.5mm} 
\scalebox{0.66}[0.66]{\includegraphics{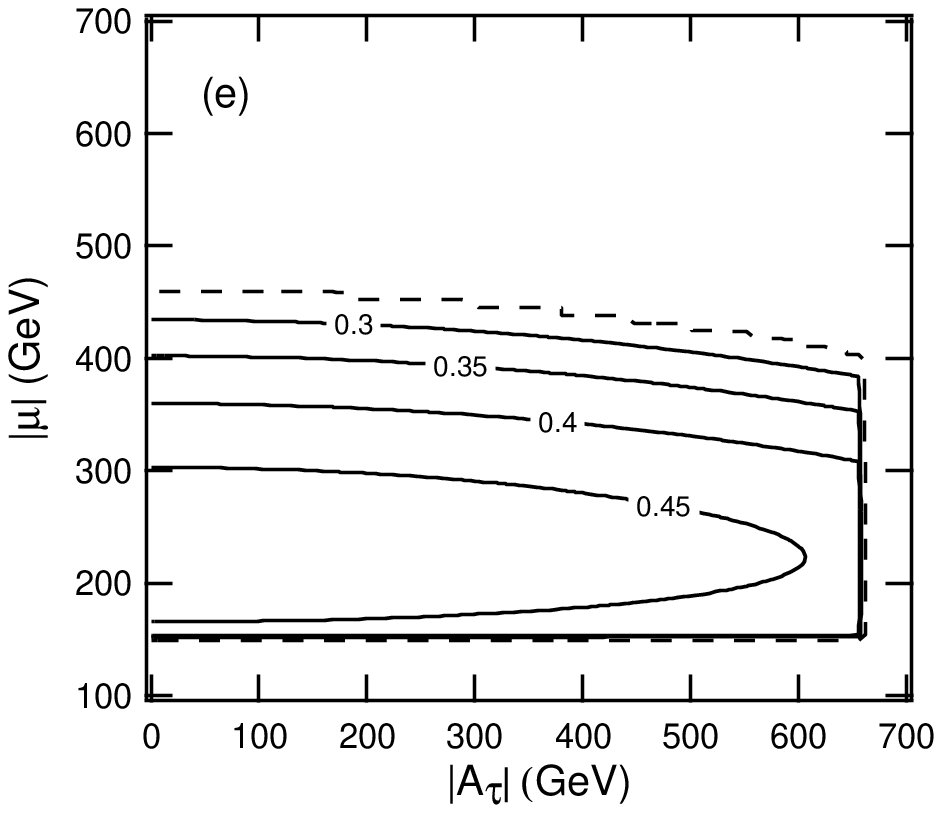}} \\ 
\scalebox{0.66}[0.66]{\includegraphics{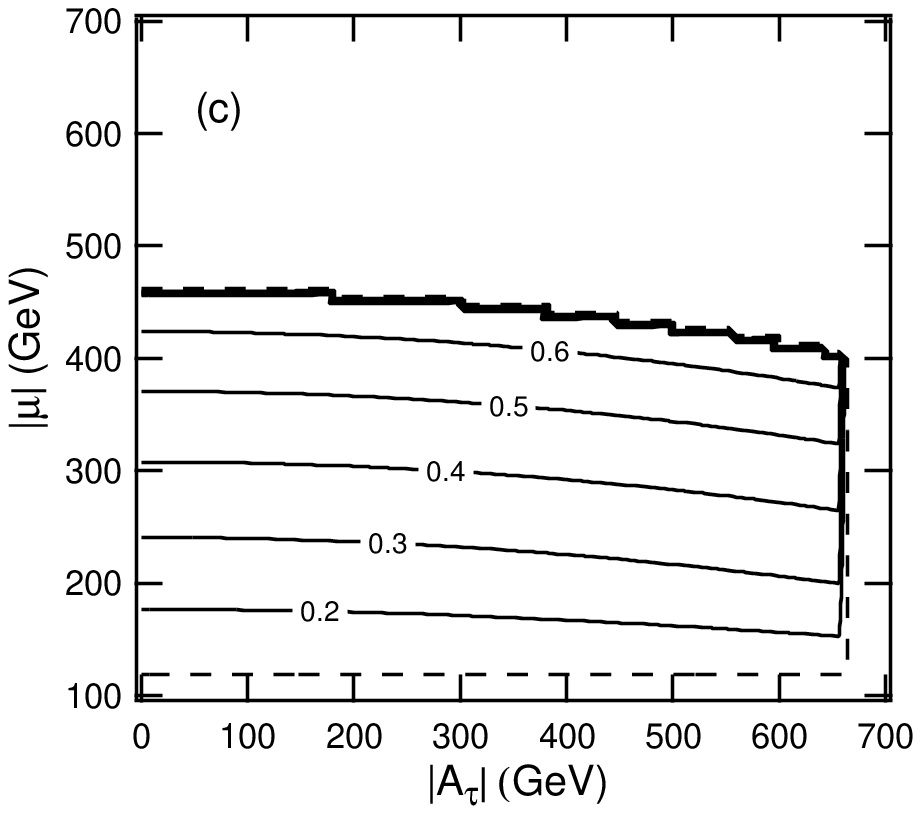}} \hspace{-7.5mm} 
\scalebox{0.66}[0.66]{\includegraphics{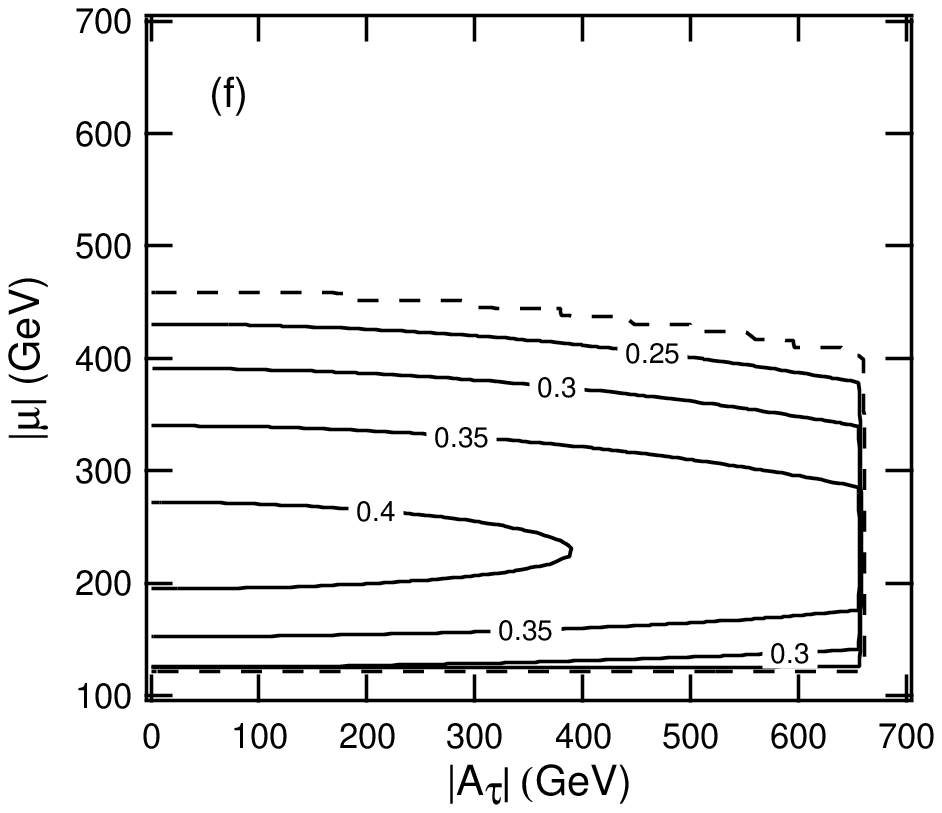}} \\ 
%
%
\vspace{5mm}
{\LARGE \bf Fig.1}
\end{center}
\end{figure}
%

\newpage
%
%
\begin{figure}[!htb] 
\begin{center}
\scalebox{0.7}[0.7]{\includegraphics{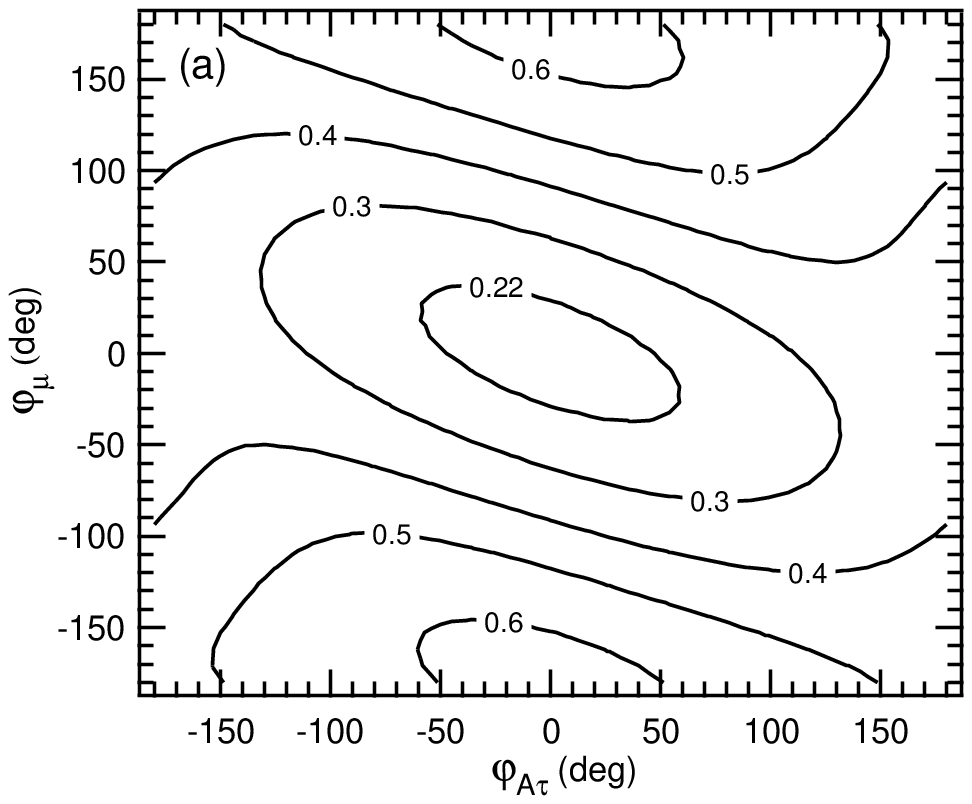}} \\ 
\scalebox{0.7}[0.7]{\includegraphics{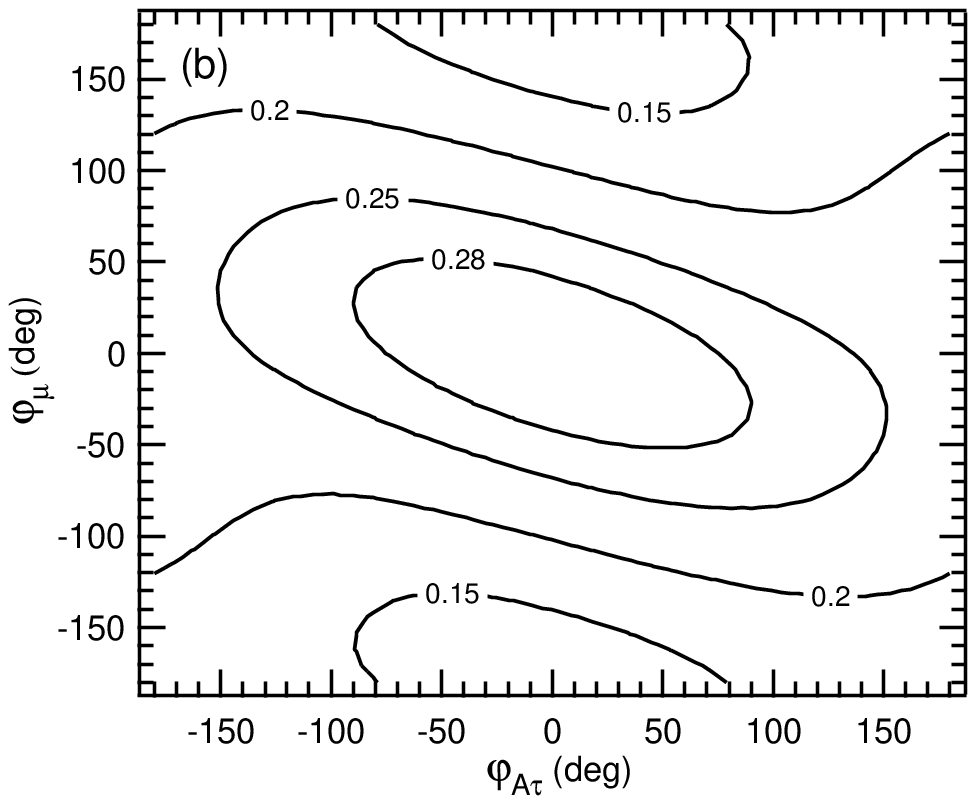}} \\ 
\scalebox{0.7}[0.7]{\includegraphics{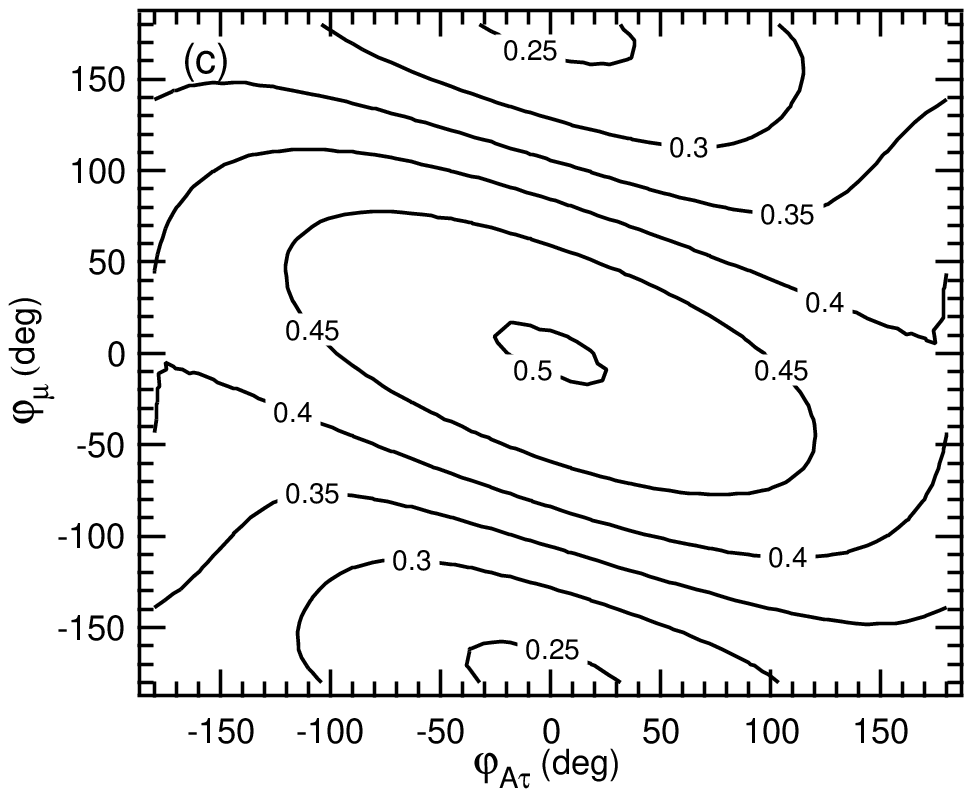}} 
\end{center}
\end{figure}

\begin{center}
{\LARGE \bf Fig.2}
\end{center}

\newpage
%
%
\begin{figure}[!htb] 
\begin{center}
\scalebox{0.8}[1.0]{\includegraphics{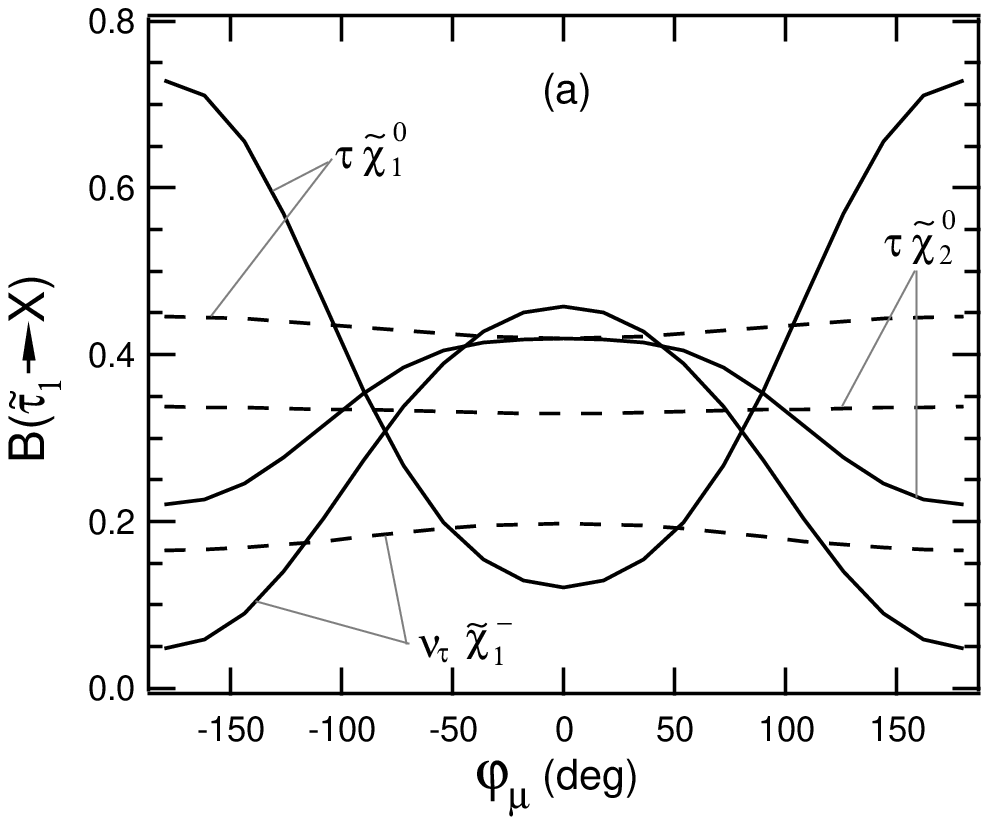}} \\ 
\scalebox{0.8}[1.0]{\includegraphics{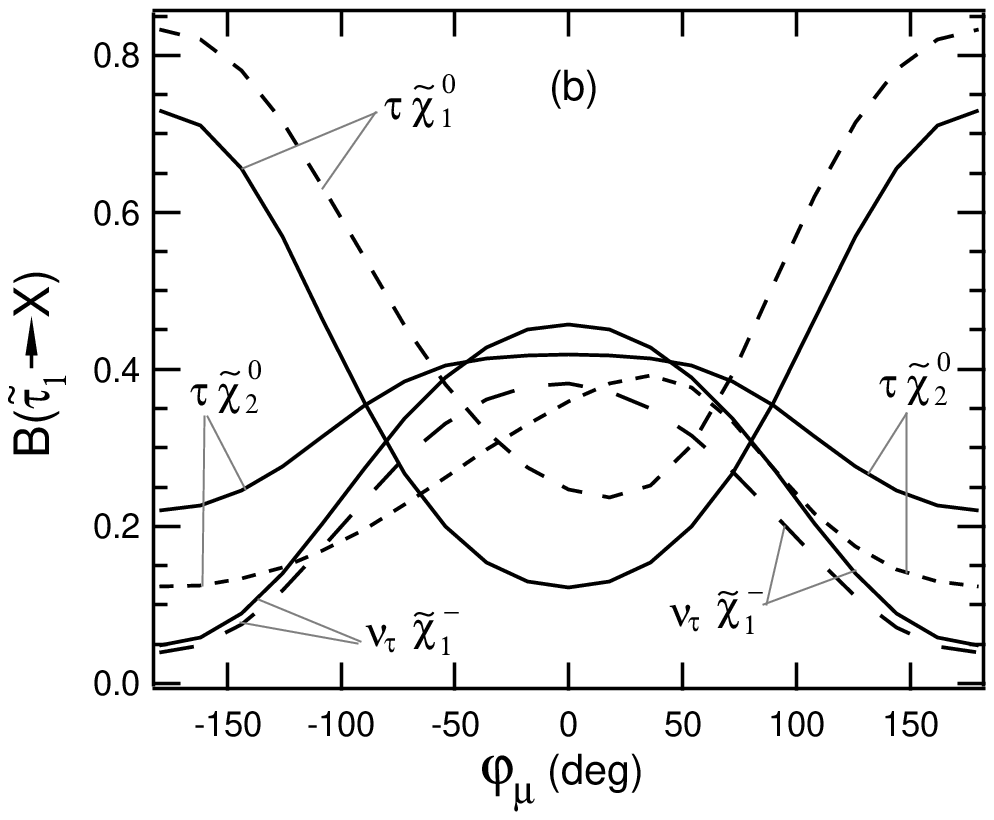}} 
\end{center}
\end{figure}

\begin{center}
{\LARGE \bf Fig.3}
\end{center}

\newpage
%
%
\begin{figure}[!htb] 
\begin{center}
\scalebox{0.7}[0.7]{\includegraphics{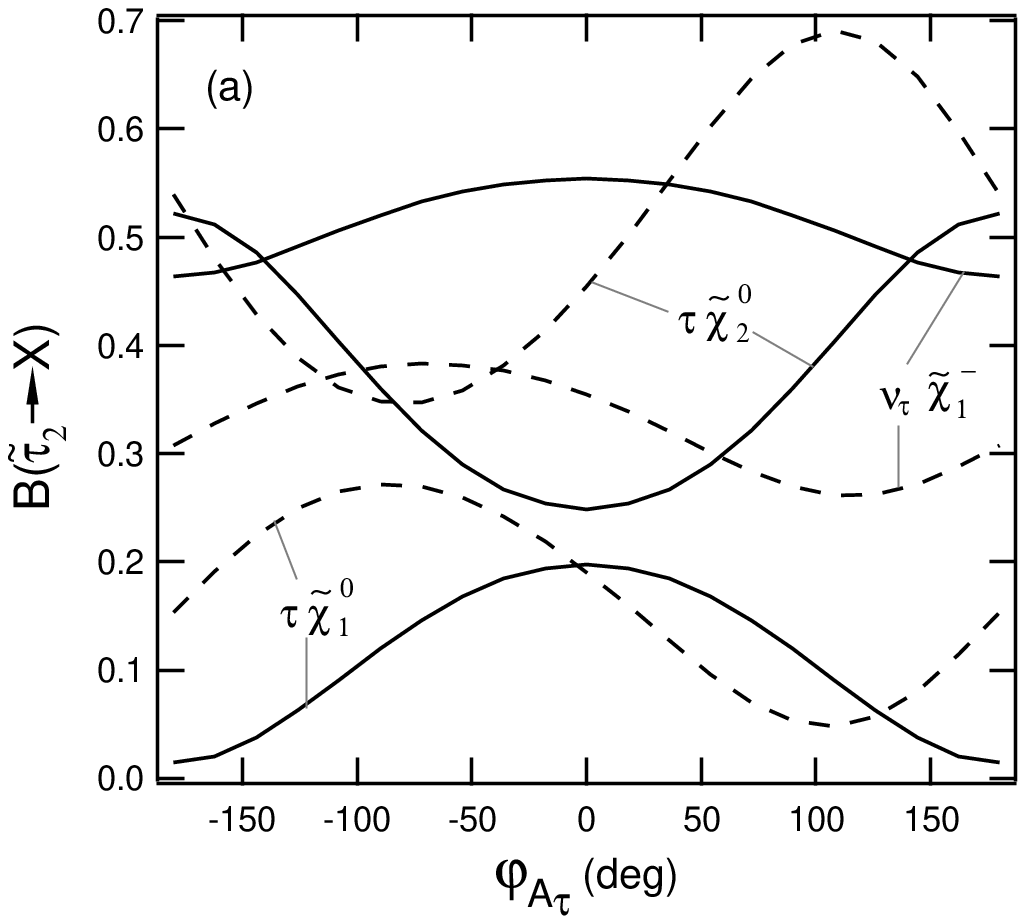}} \\ 
\vspace{-2mm}
\scalebox{0.7}[0.7]{\includegraphics{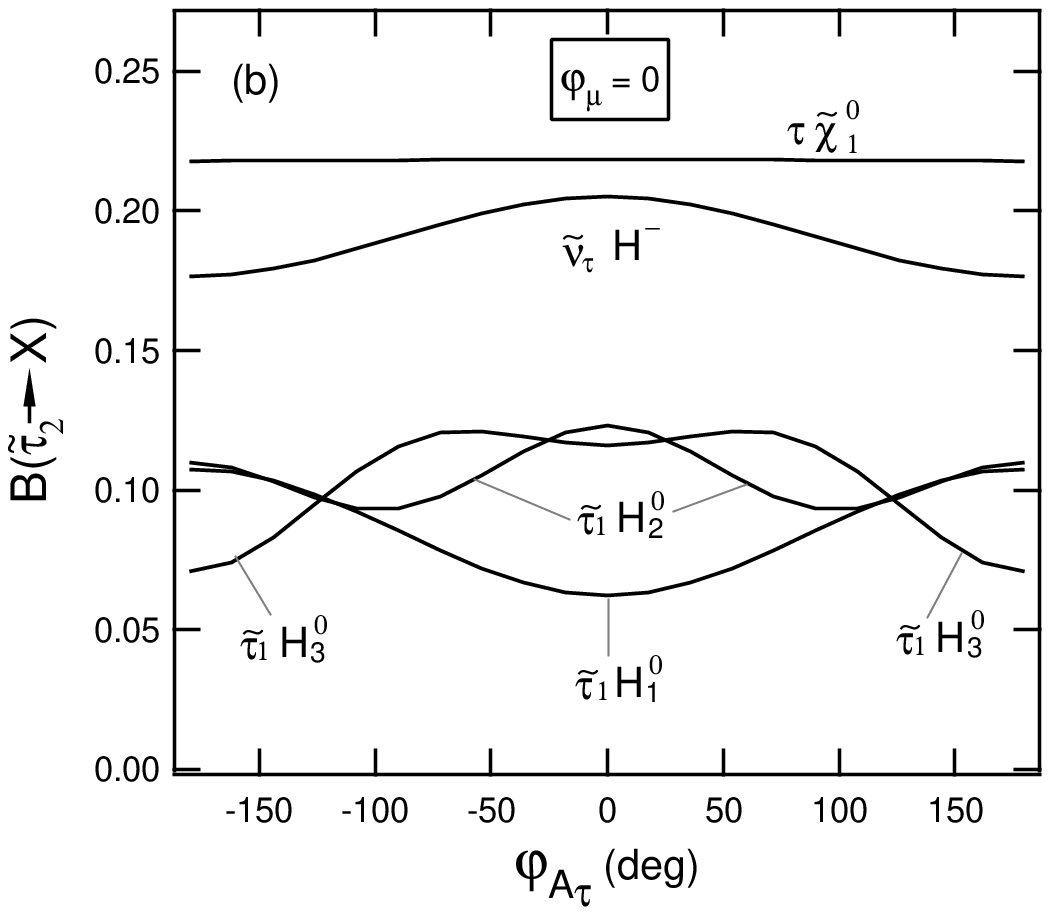}} \\ 
\vspace{-2mm}
\scalebox{0.7}[0.7]{\includegraphics{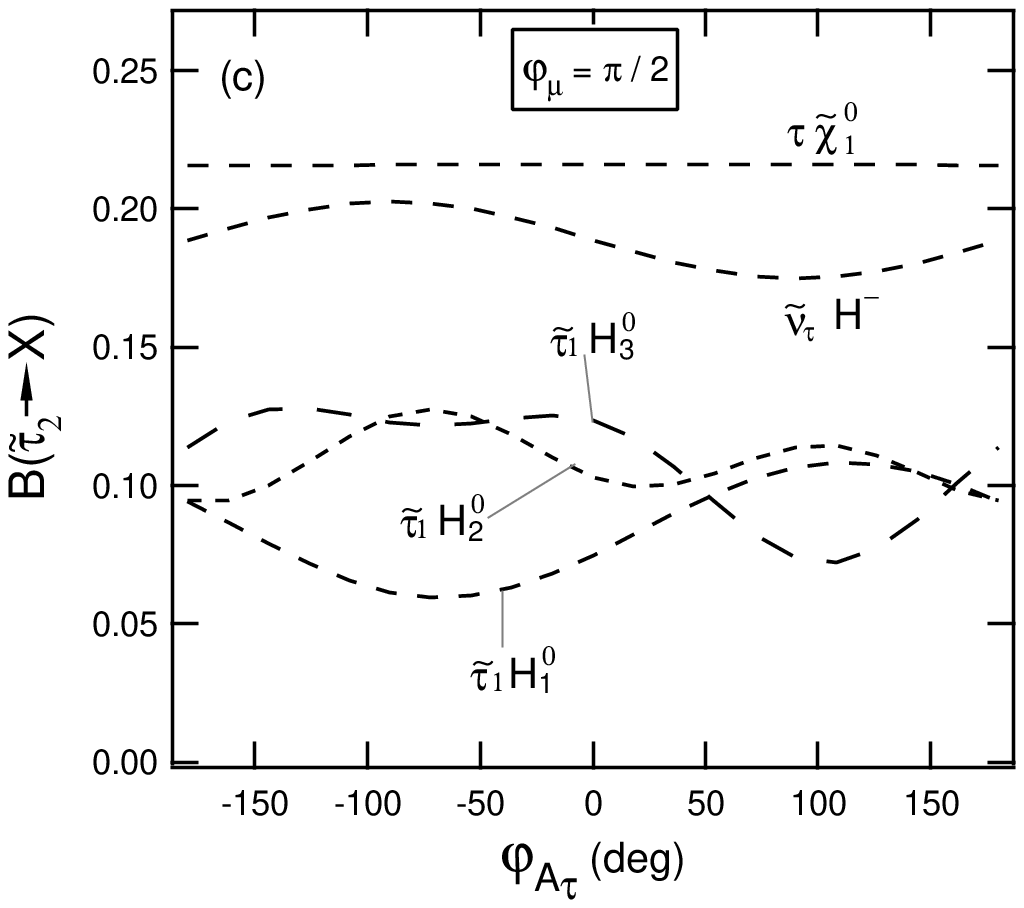}} \\ 
\vspace{5mm}
{\LARGE \bf Fig.4}
\end{center}
\end{figure}
%

\newpage
%
%
\begin{figure}[!htb] 
\begin{center}
\scalebox{1.0}[1.2]{\includegraphics{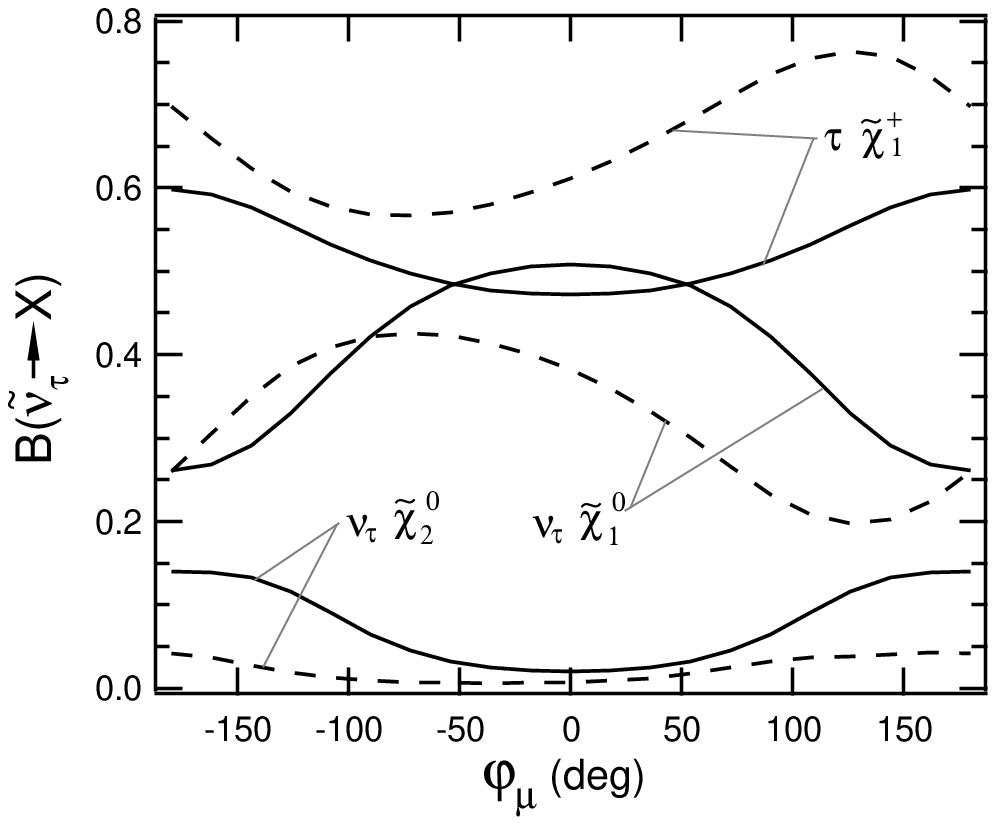}} \\ 
\vspace{5mm}
{\LARGE \bf Fig.5}
\end{center}
\end{figure}
%


\end{document}